# Selective Fabrication of Monolayer 1*H*- and 1*T'*-WTe$_2$

Ryuichi Ando[1], Katsuaki Sugawara[1,2,3,*], Tappei Kawakami[1], Takashi Takahashi[1], and Takafumi Sato[1,2,4,5,6,*]

[1] *Department of Physics, Graduate School of Science, Tohoku University, Sendai 980-8578, Japan*

[2] *Advanced Institute for Materials Research (WPI-AIMR), Tohoku University, Sendai 980-8577, Japan*

[3] *Precursory Research for Embryonic Science and Technology (PRESTO), Japan Science and Technology Agency (JST), Tokyo 102-0076, Japan*

[4] *Center for Science and Innovation in Spintronics (CSIS), Tohoku University, Sendai 980-8577, Japan*

[5] *International Center for Synchrotron Radiation Innovation Smart (SRIS), Tohoku University, Sendai 980-8577, Japan*

[6] *Mathematical Science Center for Co-creative Society (MathCCS), Tohoku University, Sendai 980-8578, Japan*

We selectively fabricated monolayers of octahedral (1*H*) and distorted trigonal (1*T'*) WTe$_2$ on graphene/SiC(0001) by controlling the substrate temperature during epitaxy. Angle-resolved photoemission spectroscopy, combined with first-principles band-structure calculations, has revealed several drastic differences between these two polymorphs. The 1*T'* phase exhibits a semiconducting character with a nearly-zero energy gap, while the 1*H* phase shows a large band gap and the band splitting at the K/K' point. The present results pave a pathway toward developing nanoelectronic devices based with WTe$_2$.

Layered transition metal dichalcogenides (TMDs) exhibit a variety of novel quantum phenomena such as superconductivity, charge-density wave, and metal-insulator transition, depending on the constituent element and the crystal symmetry.[1] TMDs generally take two different types of stable structures; one is 2*H* with a triangular prismatic structure stacked with a two-fold periodicity, and the other is 1*T* with an octahedral structure. It is also known that bulk TMDs containing the group-VI transition metal such as Mo and W take some additional crystal structures different from 2*H* and 1*T* when the growth condition is well controlled. This is highlighted by realization of 1*T'*-MoTe$_2$ upon local heating of bulk 2*H*-MoTe$_2$ crystal, leading to an ohmic homojunction with a very high mobility.[2]

Here we focus on WTe$_2$ having two stable structures in bulk, 2*H* and $T_d$.[3] The latter takes an orthorhombic structure with distorted octahedral 1$T'$ layers stacked alternately with 180° rotation.[4] Bulk 2$H$-WTe$_2$ is an indirect semiconductor[5] whereas the monolayer counterpart (1$H$) is predicted to be a direct-gap semiconductor with a spin-split band structure at the K/K' point in the Brillouin zone (BZ) due to the strong spin-orbit coupling (SOC) and the space-inversion-symmetry (SIS) breaking.[6] The SIS breaking in monolayer 1$H$ phase is known to play an essential role in realizing various exotic properties such as circular-light-polarization-dependent photoluminescence[7] and Ising superconductivity[8] in pristine monolayer and carrier-doped monolayer 1$H$-MoS$_2$, respectively. On the other hand, bulk $T_d$-WTe$_2$ is a Weyl semimetal with surface Fermi-arc states,[9,10] whereas the monolayer counterpart (1$T'$-WTe$_2$) is a candidate of two-dimensional topological insulator (2D TI),[11] as supported by the observation of edge states in the transport and scanning-tunneling-microscopy (STM) measurements.[12,13] While it is important to establish a method to selectively fabricate various crystal phases in order to further functionalize the W-based TMDs, this has yet to be established for monolayer WTe$_2$.

In this work, we have succeeded in selectively fabricating 1$H$ and 1$T'$ monolayers of WTe$_2$ by precisely controlling the substrate temperature ($T_s$) during the molecular-beam epitaxy (MBE), and characterized the band structure using ARPES and DFT (density-functional-theory) calculations.

Monolayer WTe$_2$ was grown on bilayer graphene.[14-16] At first, bilayer graphene was fabricated by resistive heating of an $n$-type 4$H$-SiC(0001) single-crystal wafer at 1100 °C for 15 min under high vacuum better than $1.0 \times 10^{-9}$ Torr. Subsequently, a monolayer WTe$_2$ film was grown by evaporating W atoms on the bilayer graphene substrate in Te-rich atmosphere. We optimized the growth condition by systematically changing $T_s$ during the epitaxial growth, and found that the optimum $T_s$ value to grow 1$H$ and 1$T'$ films is 280 °C and 310 °C, respectively [see Fig. 1(a)]. To improve the crystallinity, as-grown monolayer WTe$_2$ films were annealed for 30 min at the same $T_s$. *In-situ* ARPES measurements were carried out using an MBS-A1 electron analyzer with a He discharge lamp (photon energy $h\nu$ = 21.218 eV) at Tohoku University and a DA-30 electron analyzer (Omicron-Scienta) at beamline BL-28A in Photon Factory, KEK. The energy and angular resolutions were set to be 16 meV and 0.2°, respectively. DFT calculations were carried out by using the QUANTUM-ESPRESSO package[17] with generalized gradient approximation[18] and also with Heyd-Scuseria-Ernzerhof (HSE06) hybrid functional. The plane-wave cutoff energy and uniform $k$-point mesh were set to be 60 Ry and 10×5×1, respectively. The thickness of inserted vacuum layer was more than 10 Å to prevent interlayer interaction. SOC was included in the calculation.

We show in Fig. 1(b) the plot of ARPES intensity at $T$ = 40 K for monolayer WTe$_2$ fabricated at $T_s$ = 280 °C, measured along the ΓK cut of graphene BZ. Figure 1(d) shows the corresponding energy distribution curve (EDC) at the Γ point. One can recognize several key features such as (i) suppression of ARPES intensity in the binding-energy ($E_B$) range of Fermi level ($E_F$)-0.5 eV indicative of the semiconducting nature, (ii) a hole band topped at $E_B$ ~1.3 eV at the Γ point which rapidly moves toward higher $E_B$ on moving away from the Γ point, and (iii) broad intensity at $E_B$ ~0.5-1 eV around the K point. These key features are reproduced well in the DFT calculations including SOC for free-standing monolayer 1H-WTe$_2$ (red curves), suggesting that the monolayer fabricated at $T_s$ = 280 °C has the 1H phase. On the other hand, in the monolayer fabricated at $T_s$ = 310 °C [Fig. 1(c, e)], one can identify an obvious difference; four hole bands topped at $E_B$ ~ 0.1, 0.6, 0.8, and 1.3 eV are observed at the Γ point. They are reasonably reproduced by the DFT calculations with HSE06 (red curves) for free-standing monolayer 1T'-WTe$_2$, consistent with the previous study[15] (note that the temperature evolution of the band structure in the 1T' phase is reported in a separate work[16]). These results indicate that monolayer 1H and 1T'-WTe$_2$ phases can be selectively fabricated by fine-tuning $T_s$ during the MBE growth. This conclusion is supported by the Te 4d core-level spectrum. While each Te $4d_{3/2}$ and $4d_{5/2}$ spin-orbit satellite is composed of a single peak in the 1H phase [see Fig. 1(f)], a two-peaked structure is identified for the 1T' phase [Fig. 1(g)]. This is because the 1H phase has a single W-Te bond length whereas the 1T' phase has two types of bond lengths due to the distorted structure[19], as shown in Fig. 1(a). It is noted here that the electron diffraction experiment does not work well to distinguish the two-fold-symmetric 1T' phase because of the inevitable mixture of three domains rotated by 120° from each other due to symmetry mismatch with the graphene substrate with six-fold symmetry, as reported in 1T'-WTe$_2$[15] and other TMDs[19]. To resolve the band structure of each domain, nano-ARPES measurement is necessary.

Now we examine the band structure of 1H phase in more detail. It is theoretically predicted that the band structure of 1H phase shows a spin-valley coupling at the K/K' point [inset to Fig. 2(a)],[6,14] causing the spin splitting due to the SIS breaking and the strong SOC. We show in Fig. 2(a) the second-derivative intensity of ARPES spectrum, compared with the DFT calculation at $E_B$ = 0.4-1.7 eV where the band splitting is expected to occur around the K/K' point. In the DFT calculation, we identify the spin-split partners topped at $E_B$ = 0.7 and 1.2 eV, respectively, which look to show a reasonable agreement with the experiment despite the broad ARPES intensity. To further examine the band splitting, we show in Fig. 2(b) the energy distribution curve (EDC) at the K point. The EDC consists of a peak at ~1.5 eV accompanied by a shoulder-like feature at ~1 eV. These features are well reproduced by numerical simulations by assuming two peaks with a moderate background. This suggests the existence of a band splitting of ~0.5 eV at the K point in monolayer 1$H$-WTe$_2$.

It is emphasized here that the success of selective fabrication of 1$H$- and 1$T'$-WTe$_2$ is an important step to realize several exotic quantum phenomena as well as for device applications. Monolayer 1$T'$-WTe$_2$ would be used in 2D-TI-based devices where the band inversion and resultant spin current are controlled by an external electronic field. Monolayer 1$H$-WTe$_2$ is expected to serve as a useful platform to realize various exotic spin-valley-coupled phenomena such as the spin- or valley-Hall effect induced by circularly polarized light.[6] In addition, a lateral heterostructure consisting of 1$T'$- and 1$H$-WTe$_2$ would be used to generate chiral spin-currents at the topological edge by injecting spin-polarized carriers across the 1$H$-1$T'$ interface.

**Acknowledgment**

This work was supported by JST-CREST (no. JPMJCR18T1), JST-PRESTO (no. JPMJPR20A8), Grant-in-Aid for Scientific Research (JSPS KAKENHI Grant Numbers JP18H01821, JP20H01847, JP20H04624, JP21H01757, JP21K18888, JP21H04435, and JP22J13724), KEK-PF (Proposal No. 2020G669, 2021S2-001, and 2022G007), Foundation for Promotion of Material Science and Technology of Japan, Samco Foundation, and World Premier International Research Center, Advanced Institute for Materials Research. T. K. acknowledge support from GP-Spin at Tohoku University. T. K. also acknowledges support from JSPS.

*E-mail: k.sugawara@arpes.phys.tohoku.ac.jp, and t-sato@arpes.phys.tohoku.ac.jp

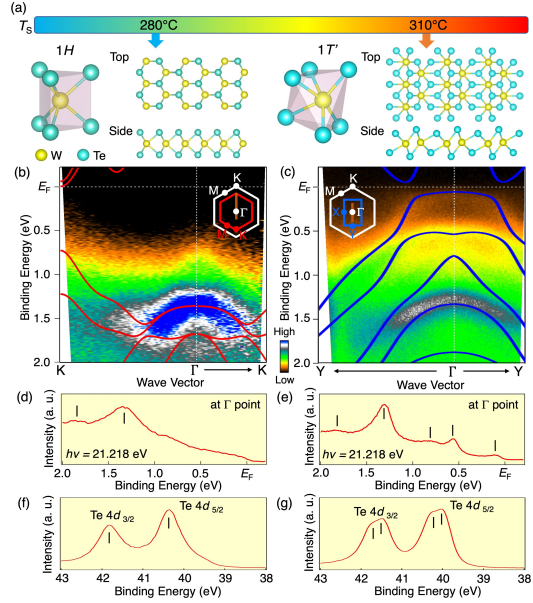

FIG. 1 (Color online). (a) Schematics of selective fabrication of 1$H$- and 1$T''$-WTe$_2$ monolayers by controlling the substrate temperature $T_s$. (b), (c) ARPES intensity plotted as a function of wave vector and binding energy for 1$H$- and 1$T''$-WTe$_2$ monolayers, respectively, measured along the ΓK cut of graphene BZ with $h\nu$ = 21.218 eV. Inset shows the first BZ of graphene (white line), 1$H$- (red line), and 1$T''$- (blue line) WTe$_2$. Corresponding DFT calculations for free-standing WTe$_2$ are overlaid for comparison. Calculations for 1$H$ and 1$T'$ phases were carried out with GGA and HSE06, respectively. (d), (e) EDC at the Γ point for 1$H$- and 1$T''$- WTe$_2$, respectively. (f), (g) Te 4$d$ core-level spectrum for 1$H$- and 1$T''$- WTe$_2$, respectively, measured at $T$ = 40 K with $h\nu$ = 80 eV. Peak positions are marked by vertical lines. While the GGA calculation overall reproduces the experimental valence-band structure for both the 1$H$ and 1$T''$ phases, it was necessary to use HSE06 for the 1$T''$ phase to reproduce the narrow band gap around $E_F$.

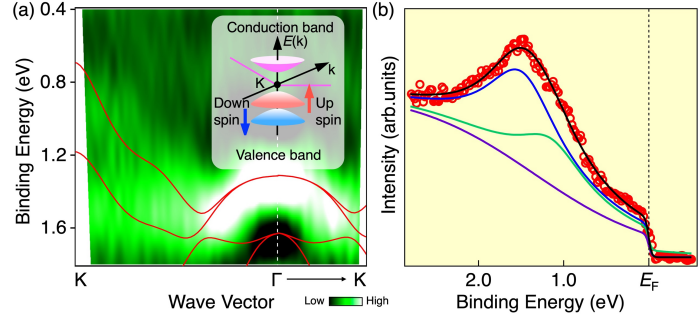

FIG. 2 (Color online). (a) Plot of second derivative intensity of ARPES spectrum for monolayer 1$H$-WTe$_2$, compared with the calculated band structure. Inset shows the schematic of spin-valley-coupled band structure around the K/K' point. (b) EDC at the K point and the result of numerical simulation (black curve) that assumes two Lorentzians (blue and green curves) with a moderate background (purple curve) multiplied by the Fermi-Dirac distribution function convoluted with the resolution function.